\documentstyle[preprint,aps,prabib]{revtex}
\begin{document}
\draft
\title{Confining solutions of $SU(3)$ Yang - Mills theory}
\author{V. Dzhunushaliev
\thanks{E-mail address: dzhun@freenet.bishkek.su}}
\address{Dept. of Physics, Virginia Commonwealth University, 
Richmond, VA 23284-2000 and Theor. Physics Dept.,
Kyrgyz State National University,
720024, Bishkek, Kyrgyzstan}
\author{D. Singleton
\thanks{E-mail address : dasingle@maxwell.phys.csufresno.edu}}
\address{Dept. of Physics, California State University Fresno,
M/S 37, Fresno, CA 93740-8031}
\date{\today}
\maketitle
\begin{abstract}
Spherically and cylindrically symmetric solutions of  $SU(3)$
Yang - Mills theory are found, whose gauge potentials
have confining properties. The spherically
symmetric solutions give field distributions which have a spherical
surface on which the gauge fields become infinite (which is
similiar to bag models of confinement), and the other solution has
a potential which increases at large distances.
The cylindrically symmetric solution 
describes a classical field ``string'' (flux tube) of the kind 
which is expected to form between quarks in the dual superconductor
picture of confinement. These solutions with classical confining
behaviour appear to be typical solutions for the classical $SU(3)$ Yang - 
Mills equations. This implies that the confining 
properties of the classical $SU(3)$ Yang - Mills theory are 
general properties of this theory.
\end{abstract}
\narrowtext

\section{Introduction}

The strong, nuclear interaction (quantum chromodynamics or QCD)
is thought to be described by a quantized SU(3) gauge theory.
In this paper we will examine solutions to the classical field
equations of an SU(3) gauge theory. The reason
for investigating these classical field configurations is to
see if they might shed some light on the confinement mechanism 
which is hypothesized to occur in the strong interaction. Although
a full explanation of the confinement mechanism may require that one
consider the fully quantized theory, the solutions presented in this 
paper have properties which mimick the behaviour of various phenomenological
explanations of confinement. In particular the various solutions
exhibit a bag-like structure similiar to the bag models \cite{bag} 
of confinement, an almost linearly increasing potential such as those 
used in the study of heavy quark bound states \cite{eich}, and a
string like structure as found in the dual superconducting picture of
confinement. 
  
The draw back of the classical configurations presented here is that they
all have infinite field energy when their energy densities are intergrated
over. This can be compared with the finite energy monopole 
and dyon solutions of Yang-Mills field theory \cite{thooft} \cite{bps}. 
At the classical level one might expect that only solutions which have 
fields that become infinite (and thus have an infinite field energy) are 
capable of giving a confining behaviour. In the context of $SU(2)$ 
Yang-Mills theory it has been shown \cite{swank} that, at the classical
level, finite energy solutions, like monopoles, do 
not lead to confinement, while infinite energy solutions do lead to 
confinment. Quantum effects may modify these classical solutions 
to soften the infinite field strengths and energies in the same
way that quantum effects soften the singularity of the Coulomb solution
in E\&M.

\section{Spherically symmetric ansatz}

The ansatz for the $SU(3)$ gauge field we take as in \cite{palla}
\cite{gal}:
\begin{eqnarray}
A _0 & = & \frac{2\varphi(r)}{{\bf i} r^2} \left( \lambda ^2 x - \lambda ^5 y 
      + \lambda ^7 z\right ) + \frac{1}{2}\lambda ^a
      \left( \lambda ^a _{ij} + \lambda ^a_{ji} \right ) 
      \frac{x^ix^j}{r^2} w(r),
\label{1a:1}\\
A^a_i & = & \left( \lambda ^a_{ij} - \lambda ^a_{ji} \right )
        \frac {x^j}{{\bf i} r^2} \left(f(r) - 1\right ) +
        \lambda ^a_{jk} \left (\epsilon _{ilj} x^k + 
	\epsilon _{ilk} x^j\right ) \frac{x^l}{r^3} v(r)
\label{1b:2},
\end{eqnarray}
here $\lambda ^a$ are the Gell - Mann matrices; $a=1,2,\ldots ,8$
is a color index; the Latin indices $i,j,k,l=1,2,3$ are the space indices; 
${\bf i}^2=-1$; $r, \theta, \varphi$ are the usual spherically coordinates. 
Substituting Eqs. (\ref{1a:1}) - (\ref{1b:2}) into the Yang - Mills equations 
\begin{equation}
\frac{1}{\sqrt{-g}}\partial _{\mu} \left (\sqrt {-g} {F^{a\mu}}_{\nu}
\right ) + f^{abc} {F^{b\mu}}_{\nu} A^c_{\mu} = 0,
\label{2}
\end{equation}
gives the following system of equations for $f(r), v(r), w(r)$ 
and $\varphi (r)$ 
\begin{eqnarray}
r^2f''& =& f^3 - f + 7fv^2 + 2vw\varphi - f\left (w^2 + \varphi ^2\right ),
\label{3a:1}\\
r^2v''& = & v^3 - v + 7vf^2 + 2fw\varphi - v\left (w^2 + \varphi ^2\right ),
\label{3b:2}\\
r^2w''& = & 6w\left (f^2 + v^2\right ) - 12fv\varphi,
\label{3c:3}\\
r^2\varphi''& = & 2\varphi\left (f^2 + v^2\right ) - 4fvw.
\label{3d:4}
\end{eqnarray}
This set of equations is difficult to solve even numerically
thus we will investigate various simplified cases when only
two of the functions are nonzero. Under this assumption there are 
three cases. In the first case $(f,w=0)$ or $(v,w=0)$ Eqs. 
(\ref{3a:1}) - (\ref{3d:4}) reduce to a form similiar to the system of equations
studied in \cite{julia} which yield the well known dyon solutions. 
We will examine the cases where $w = \varphi = 0$, and $f = \varphi =0$ 
(or $v=\varphi =0$).

\subsection{The $SU(3)$ bag}

In this case we set $w=\varphi=0$ so that Eqs. (\ref{3a:1}) - (\ref{3d:4}) 
reduce to the following form
\begin{eqnarray}
r^2f''& = & f^3 - f + 7fv^2,
\label{4c:3}\\
r^2v''& = & v^3 - v + 7vf^2.
\label{4d:4}
\end{eqnarray}
To simplify the equations further we take $f(r) = v(r) = q(r) / \sqrt{8}$.
This reduces Eqs. (\ref{4c:3}) - (\ref{4d:4}) to
\begin{equation}
\label{5}
r^2 q '' = q(q^2 -1)
\end{equation}
This is the Wu-Yang equation. In addition to the monopole solutions
to this equation \cite{wy} \cite{pagel} it is also known that 
this equation possesses a solution which becomes infinite on a spherical
surface \cite{swank} \cite{rosen} \cite{adl} \cite{per} \cite{lun}. If 
one lets this spherical surface be at $r=r_0$ then in the limit 
$r \rightarrow r_0$ the form of the solution approaches
\begin{equation}
\label{6}
q(r) \approx {\sqrt{2} r_0 \over r_0 - r}
\end{equation}
Using Eq. (\ref{6}) to find $f(r), v(r)$ and inserting these back into
Eq. (\ref{1b:2}) shows that the $A_i ^a$ gauge field develops a singularity
on the sphere of radius $r=r_0$. It is easy to solve Eq. (\ref{5}) 
numerically (for this work we used the {\it Mathematica} \cite{math} 
numerical differential solver routinue). In solving Eq. (\ref{5}) 
we considered that near $r=0$ the function $q(r)$ had a series 
expansion like
\begin{equation}
\label{7}
q(r) = 1 + q_2 \frac{r^2}{2 !} + ...
\end{equation}
where $q_2$ is some constant. Chosing a specific $q_2$ at some
radius $r=r_i$ sets the initial conditions on $q(r_i) , q'(r_i)$
for the numerical solution. The choice of these initial conditions
determined the radius on which $q(r)$ became singular. In Fig. 1
we show a typical example of $q(r)$. This type of field configuration is
somewhat similiar to a bag-like structure, and it has been shown that
such structures lead to the confinement of a test particle placed in
the field of this solution \cite{swank} \cite{lun} \cite{mah}
\cite{lun2}. If complex gauge
fields are allowed \cite{prot} or if scalar fields are introduced
into the field equations \cite{sing} it is possible to find analytical
solutions which possesses gauge fields which are singular on some
spherical surface of radius $r=r_0$. Several authors have remarked on
the mathematical similiarity between the above solution and the 
Schwarzschild solution of general relativity, which leads to a
gravitational type of confinement.

\subsection{The $SU(3)$ bunker}

Here we examine the $f=\varphi=0$ case. The case $v=\varphi=0$ is entirely
analogous. From Eqs. (\ref{3a:1}) - (\ref{3d:4}) the equations for the ansatz functions
become
\begin{eqnarray}
r^2 v''& = & v^3 - v - vw^2,
\label{9:1}\\
r^2 w''& = & 6w v^2.
\label{9:2}
\end{eqnarray}
Near $r=0$ we took the series expansion form for $v$ and $w$ as
\begin{eqnarray}
v = 1 + v_2 \frac{r^2}{2 !} + ...,
\label{10:1} \\
w = w_3 \frac{r^3}{3 !} + ...
\label{10:2}
\end{eqnarray}
where $v_2 , w_3$ were constants which determined the initial conditions
on $v$ and $w$ as in the last section.
In the asymptotic limit $r \rightarrow \infty$ the form of the solutions
to Eqs. (\ref{9:1}) - (\ref{9:2}) approaches the form
\begin{eqnarray}
v & \approx & A \sin \left (x^{\alpha } + \phi _0\right ),
\label{11:1}\\
w & \approx & \pm\left [ \alpha  x^{ \alpha } + 
\frac{\alpha -1 }{4}\frac{\cos {\left (2x^{\alpha} + 2\phi _0 \right )}}
{x^{\alpha}}\right ],
\label{11:2}\\
3A^2 & = & \alpha(\alpha - 1).
\label{11:3}
\end{eqnarray}
where $x=r/r_0$ is a dimensionless radius and $r_0, \phi _0$, and $A$ are 
constants. The second, strongly oscillating term in $w(r)$ is kept since
it contributes to the asymptotic behaviour of $w''$. As in the previous 
case we did not find an analytical solution for
Eqs. (\ref{9:1}) - (\ref{9:2}) but it is straight
forward to solve these equations numerically. A typical solution is shown 
in Fig. 2. The strongly oscillating behaviour of $v(r)$ resulted in the
space part of the gauge field of Eq. (\ref{1b:2}) being strongly oscillating.
The ansatz function $w(r)$ increases as some power of $x$ as $x \rightarrow
\infty$, and would lead to the confinement of a test particle placed
in the background field of this solution.
(For the bunker solution there is some subtlety
associated with the confinement of the test particle due to
pair creation when the test particle scatters off the
potential. This is essentially related to the Klein
paradox and is discussed in Refs. \cite{ram} \cite{tez}).
The type of confinement given by this bunker
solution is different from the bag-like solution of the
previous sub-section : First the confining behaviour of the bag-like
solution came from the ``magnetic'' part of the gauge field ($A_i ^a$) through
the ansatz functions $v(r), f(r)$, while in the present case it is 
the ``electric'' part of the gauge field ($A_0 ^a$) which gives confinement
through the ansatz function $w(r)$. Second, the bag-like solution
confines a test particle by the field strength becoming infinitely
large at some finite value of $r$, while the present solution confines
a test particle by the field strength increasing without bound as
$r \rightarrow \infty$. The power
law with which $w(r)$ increases changes as $r$ increases. In Fig. 3
we show a plot of $Log (w) - Log (x)$ for the solution of Fig. 2. At 
around $Log (x) \approx 0.7$ the slope of the line (and therefore the power
law increase of $w(r)$) changes from $\alpha \approx 2.8$ to $\alpha
\approx 1.3$. Depending on the initial conditions we found that for
$x$ near the origin $\alpha$ was in the range $\approx 2 -3$ while
as $x$ became large $\alpha$ decreased to the range of $\approx 1.2 -
1.8$. In studies of heavy quark bound states \cite{eich} a potential
which increases as $r \rightarrow \infty$ is often used to
sucessfully model the excited states of these systems. In these studies 
the increase is usually linear in $r$.
  
The ``magnetic'' and ``electric'' fields associated with this solution
can be found from $A_{\mu} ^a$, and have the following behaviour 
\begin{eqnarray}
\label{12:1}
H^a _r & \propto & \frac{v^2-1}{r^2} , \; \; \; \; \; \; \; \;
H^a_{\varphi}  \propto  v' , \; \; \; \; \; \; \; \;
H^a_{\theta}  \propto  v' , \\
\label{12:2}
E^a_r & \propto & \frac{rw' - w}{r^2}, \; \; \; \; \; \; \; \;
E^a_{\varphi} \propto  \frac{vw}{r}, \; \; \; \; \; \; \; \; 
E^a_{\theta}  \propto  \frac{vw}{r},
\end{eqnarray}
here for $E^a _r , H^a _{\theta}$, and $H^a_{\varphi}$ the color index 
$a=1,3,4,6,8$ and for $H^a _r, E^a _{\theta}$ and $E^a _{\varphi}$ 
$a=2,5,7$. The asymptotic behaviour of 
$H^a_{\varphi}, H^a_{\theta}$ and $E^a_{\varphi}, E^a_{\theta}$ 
is dominated by the strongly oscillating function $v(r)$. If 
quantum corrections where applied to this solution it is
expected that these strongly oscillating fields would
be smoothed out and not play a significant role in the large $r$
limit. From Eqs. (\ref{12:1}) - (\ref{12:2}) and the asymptotic form of $v(r), w(r)$
the radial components of the ``magnetic'' and ``electric'' 
have the following asymptotic behaviour
\begin{equation}
\label{13}
H^a_r  \propto  \frac{1}{r^2}, \; \; \; \; \; \;
E^a_r  \propto  \frac{1}{r^{2-\alpha}}.
\end{equation}
where the strongly oscillating portion of $H^a _r$ is assumed not
to contribute in the limit of large $r$ due to smoothing by quantum
corrections. The radial ``electric'' field falls off slower than 
$1/r^2$ (since $\alpha > 1$) indicating the presence of a confining
potential. The $1/r^2$ fall off of $H^a _r$ indicates that this solution
carries a ``magnetic'' charge. This was also true for the simple solutions
discussed in Refs. \cite{palla} \cite{pagel}. It can also be shown in
the same way that the bag-like solution of the previous section also
carries a ``magnetic'' charge. This leads to the result
that if a test particle
is placed in the background field of either the bag or bunker solution,
this composite system will have unusal spin properties \cite{rebbi}
({\it i.e.} if the test particle is a boson the system will behave
as a fermion, and if the test particle is a fermion the system will
behave as a boson).

Just as for the bag solution, the biggest
draw back of the present solution is its infinite
field energy. The bunker solution has an
asymptotic energy density proportional to
\begin{equation}
{\cal E} \propto 4\frac{v'^2}{r^2} + \frac{2}{3}\left (
\frac{w'}{r} - \frac{w}{r^2}\right ) ^2 + 4\frac{v^2w^2}{r^4} +
\frac{2}{r^4} \left (v^2 - 1\right )^2 \approx \frac{2}{3}
\frac{\alpha ^2 (\alpha -1) (3 \alpha -1)}{x^{4-2\alpha}}
\label{14}
\end{equation}
Since we found $\alpha > 1$ this energy density will yield an
infinite field energy when integrated over all space. This can
be compared with the finite field energy monopole and dyon solution
\cite{bps}. However, as remarked previously, it has been
demonstrated \cite{swank} that the finite energy monopole solutions
do not trap a test particle while the infinite energy solutions
do.

What is the physical meaning of this solution ? As in the case of
the bag-like solution one can examine the motion of a test
particle in the background field of the bunker solution, and
find in this way that the test particle will tend to remain
confined due to the increasing gauge potential.
Another possible interpretation is that this solution is
the Yang-Mills analog to the Coulomb potential in electrostatics. 
An electron can exist as an asymptotic state while a quark can not.
Therefore, the bunker solution can be  
thought of as the far field of a color charge - ``quark''. The fact
that the bunker solution possesses an infinite field energy then indicates
that an isolated quark is not allowed as an observable free state. The
Coulomb solution of electrostatics also posseses an infinite field
energy, but the manner in which the field energy becomes infinite
is different than for the bunker solution. Any point electric charge
such as the electron has a singularity at $r=0$, but  
the ``quark'' field of the bunker solution has a singularity at
$r=\infty$. To follow through on
this interpretation of the bunker solution as an isolated
``quark'', one should investigate
what happens when two bunker solutions are placed in the
vicinity of one another. In this way one might hope that
the combination of two bunker solutions would lead to
a localized, finite energy field configuration. Then
if one tried to separate the two ``quarks'' the field energy
would become infinite. However the nonlinear character
of the classical SU(3) field equations make this a difficult
problem beyond the scope of the present work. Finally it can be
noted that this solution is in a sense asymptotical free since
at $r=0$ the gauge potential $A^a_{\mu} \to 0$.

\section{The gauge ``string''}

Let us write down the following ansatz 
\begin{eqnarray}
A^2_t & = & f(\rho ),
\label{16:1}\\
A^5_z & = & v(\rho ),
\label{16:2}\\
A^7_{\varphi} & = & \rho w(\rho ),
\label{16:3}
\end{eqnarray}
here we use the cylindrical coordinate system $z, \rho , \varphi$.
The color index $a=2,5,7$ corresponds to an embedding of  $SU(2)$
in $SU(3)$. Using Eqs. (\ref{16:1}) - (\ref{16:3}) the Yang - Mills equations become
\begin{eqnarray}
f'' + \frac{f'}{\rho} & = & f\left (v^2 + w^2 \right ),
\label{17:1}\\
v'' + \frac{v'}{\rho} & = & v\left (-f^2 + w^2 \right ),
\label{17:2}\\
w'' + \frac{w'}{\rho}  - \frac{w}{\rho ^2}& = & w
\left (-f^2 + v^2 \right ),
\label{17:3}
\end{eqnarray}
Let us examine the simple case $w=0$ which reduces Eqs.
(\ref{17:1}) - (\ref{17:3}) to
\begin{eqnarray}
f'' + \frac{f'}{\rho} & = & fv^2 ,
\label{18:1}\\
v'' + \frac{v'}{\rho} & = & -vf^2 .
\label{18:2}
\end{eqnarray}
At origin $\rho =0$ the solution has the following form
\begin{eqnarray}
f & = & f_0 + f_2\frac{\rho ^2}{2} + \ldots ,
\label{19:1}\\
v & = & v_0 + v_2\frac{\rho ^2}{2} + \ldots .
\label{19:2}
\end{eqnarray}
Substituting Eqs. (\ref{19:1}) - (\ref{19:2})
into (\ref{18:1}) - (\ref{18:2}) we find that
\begin{eqnarray}
f_2 & = & \frac{1}{2}f_0v_0^2 ,
\label{20:1}\\
v_2 & = & -\frac{1}{2}v_0f_0^2 .
\label{20:2}
\end{eqnarray}
The asymptotic behaviour of the ansatz functions $f, v$ and the energy 
density ${\cal E}$ can be given as
\begin{eqnarray}
f & \approx & 2\left [x + \frac{\cos \left (2x^2 + 2\phi _1\right )}
{16x^3} \right ] ,
\label{21:1}\\
v & \approx & \sqrt{2} \frac{\sin \left (x^2 + \phi _1 \right )}{x} ,
\label{21:2}\\
{\cal E} & \propto & f'^2 + v'^2 + f^2v^2 \approx const,
\label{21:3}
\end{eqnarray}
where $x=\rho /\rho _0$ is the dimensionless radius, and $\rho _0, \phi _1$ 
are  constants. To solve the system in Eqs. (\ref{18:1}) - (\ref{18:2}) for all $r$
we again used numerical methods. A typical solution for $f$ and $v$
is shown in Fig. 4. As in the solution of the previous section
we have a confining potential $A^2_t=f(\rho)$ and a strongly oscillating 
potential $A^5_z=v(\rho)$. Depending on the relationship between $v_0$ and
$f_0$ the energy density near $\rho = 0$ will be either a hollow ({\it i.e.}
an energy density less than the asymptotic value) or a hump ({\it i.e.}
an energy density greater than the asymptotic value). 
On account of this and the cylindrical symmetry of this solution 
we call this the ``string'' solution. The quotation marks 
indicate that this is a string from an energetic point of view, not from 
the potential ($A^a_{\mu}$) or field strength ($F^a_{\mu\nu}$) point of view.
After quantization the oscillating functions will most likely vanish 
and only the confining potential and constant energy density will remain.

This ``string''-like  solution can be thought of as describing the 
classical gauge field between two ``quarks''. Similiar string-like 
configurations are thought to occur in the dual superconductor picture
of confinement, and lattice calculations (nonperturbative
quantization) also may give evidence for such structures.

\section{Discussion}

In this paper we have examined several non-trivial classical solutions 
of the $SU(3)$ Yang - Mills theory. Each of these solutions 
demonstrated some type of confining behaviour, indicating that this 
may be a general property of the classical $SU(3)$ Yang - Mills theory,
and also that some form of this behaviour may carry over to the
quantized theory. These infinite energy solutions to Eqs. (\ref{3a:1})
- (\ref{3d:4})
represent typical solutions to the classical field equations
in the sense that they arise for a wide range of initial conditions. 
In contrast to this the simple SU(3) monopole and dyon solutions 
investigated in Refs. \cite{palla} \cite{pagel}
are unique solutions in the sense that they arise for only certain
initial conditions. In addition the infinite energy solutions
investigated here give rise to a classical type of confining
behaviour which neither the SU(3) solutions of Refs. \cite{palla}
\cite{pagel} or the finite energy \cite{thooft} \cite{bps} solutions possess. 

The physical significance of the spherically symmetric cases is motivated
by noting the similiarities between these solutions and various
phenomenological models of confinement. The first solution
will confine a quantum test particle via the spherical singularity in
the ``magnetic'' part of the gauge field in a manner similiar to
some bag models. Studies of such bag-like field configurations
with scalar \cite{lun} and spinor \cite{mah}
\cite{lun2} test particles have been carried out. In both cases it
was found that the test particles were confined inside $r = r_0$, and in
Ref. \cite{lun2} a somewhat realistic spectrum of hadron masses
was obtained in this way. The second solution has the ``electric'' part of
the gauge field increasing like $r^{\alpha}$ for large $r$ with
$\alpha >1$. If this field configuration is taken as representing
the far field of an isolated ``quark'' then the infinitely increasing
field strength can be taken to indicate the impossibility of
isolating an individual ``quark''. In contrast
isolated electrons exist in nature since they generate 
electric fields which decrease at infinity. The third solution
has a string-like structure from an energetic point of view.
Similiar string-like structures are found in the dual superconductor
picture of confinement. Just as two interacted electrons generate 
an electric field which is essentially the superposition 
of the electric fields of the individual electrons, so two interacting 
quarks are thought to generate a string-like flux tube which
runs from one quark to the other. The ``string'' solution 
obtained above is a classical model of such a field 
distribution. It appears as a string-like structure on the 
background of the field with constant energy density. The strongly
oscillating components of this and the bunker solution will most 
likely be smoothed out once quantum effects are taken into account.

\section{Acknowledgements} This work has been funded in part by the
National Research Council under the Collaboration in Basic Science and
Engineering Program. The mention of trade names or commercial
products does not imply endorsement by the NRC.

\newpage
\centerline{{\bf List of figure captions}}

Fig.1. The $q(r)$ function for the $SU(3)$ bag. The initial conditions
for this solution were $q_2 = 0.1$ and $r_i = 0.001$. 

\vspace{1.0in}

Fig.2. The $w(x)$ confining function, and the $v(x)$ oscillating function 
of the $SU(3)$ bunker solution. The initial conditions for this 
particular solution were $v_2 = 0.1$, $w_3 = 2.0$, and $x_i = 0.001$.

\vspace{1.0in}

Fig.3. A plot of $Log(w) - Log(x)$ of the solution from Fig. 2 showing 
the different power law behaviour in the small $x$ and large $x$ regions.

\vspace{1.0in}

Fig.4. The $SU(3)$ ``string'' solution with the linearly confining
function $f(x)$ and the strongly oscillating function $v(x)$. The initial
conditions for this solution were $f_0 = 0.75$, $v_0 = 0.75$
and $x_i = 0.001$.

\end{document}